\documentclass{article}

\usepackage{PRIMEarxiv}

\usepackage{float}
\usepackage{amsmath}
\usepackage{amssymb}
\usepackage{subcaption}
\usepackage{listings}

\usepackage[utf8]{inputenc} 
\usepackage[T1]{fontenc}    
\usepackage{hyperref}       
\usepackage{url}            
\usepackage{booktabs}       
\usepackage{amsfonts}       
\usepackage{nicefrac}       
\usepackage{microtype}      
\usepackage{lipsum}
\usepackage{fancyhdr}       
\usepackage{graphicx}       
\graphicspath{{media/}}     

\title{PyelogP: Automated Energy-Based Determination of Preconsolidation Pressure in Clay Deposits
}

\author{
  Liang Chern Chow\\
  Independent Researcher \\
  Oakland, California\\
  \texttt{liangchern@gmail.com} \\
   \And
  Thierno Kane \\
  Independent Researcher \\
  Chicago, Illinois\\
  \texttt{tkane83@gmail.com} \\
}

\begin{document}
\maketitle

\begin{abstract}
Estimating the preconsolidation pressure ($\sigma'_p$) from one-dimensional consolidation (oedometer) tests is critical in geotechnical engineering for settlement analysis. Traditional graphical methods, such as the Casagrande procedure, may introduce uncertainties, particularly when interpreting rounded $e$-log($P$) curves typical of disturbed specimens of soft clays and silt deposits. This paper introduces \texttt{PyelogP}, an open-source Python library designed to calculate $\sigma'_p$ using the strain-energy method proposed by Becker et al. (1987) as an automated and reproducible alternative. The algorithm combines natural cubic spline interpolation, knee-point detection via the Kneedle algorithm, and split-point linear regression within the work-pressure space. Physically informed thresholds, including overconsolidation ratio limits and second-derivative maxima (${d^2 e}/{d(\log \sigma')^2}$), are incorporated to establish pre-yield and post-yield fitting boundaries. The performance of \texttt{PyelogP} is evaluated against a suite of 22 experimental consolidation datasets covering various clay deposits, including Saint-Alban clay and San Francisco Old Bay Clay. The results demonstrate strong agreement with the published values ($R^2$ = 0.912, RMSE = 0.374, MBE = -0.080), while the $O(N^2)$ algorithm requires only a few milliseconds per curve for typical oedometer datasets and less than 150 milliseconds for the largest datasets.
\end{abstract}

\keywords{Consolidation \and Geotechnical \and Kneedle \and Preconsolidaion Pressure \and Settlement}

\section{Introduction}
In geotechnical engineering, a settlement analysis is performed to estimate the vertical downward movement of a structure or foundation resulting from loading of the underlying soil. It is a critical component of geotechnical design and construction projects because excessive total settlements or differential settlements can lead to damage in buildings, bridges, roads, and other infrastructures. Settlement occurs due to changes in soil volume caused by changes in stresses exerted on the ground. The magnitude of the volume change depends on the compressibility of the soil. In general, clayey soils are more compressible than granular soils. To evaluate the compressibility characteristics of a soil, a one-dimensional consolidation (oedometer) test is performed on laboratory specimen on representative samples collected from the field. The primary outputs of a consolidation test include the time-deformation curves, a void ratio (deformation or vertical strain) versus effective vertical stress ($e$-log($P$)) curve, and consolidation parameters derived from these results.

One of the key parameters obtained from the $e$-log($P$) curve is the preconsolidation pressure ($\sigma'_p$). The preconsolidation pressure is a unique property of a soil that represents the maximum effective vertical stress the soil has experienced throughout its geological history. The geologic preconsolidation pressure may result from snow or ice load that has melted, thick sediment overburden that has eroded away, changing groundwater level conditions, desiccation, aging etc. The $\sigma'_p$ is commonly normalized with the in situ effective vertical stress ($\sigma'_{v0}$) and referred to as the overconsolidation ratio (OCR). Normally consolidated (NC) soils have an OCR of approximately 1.0 and are typically associated with young deposits (i.e., no geological history such that $\sigma'_p$ is approximately equal to $\sigma'_{v0}$). In contrast, most natural clay and silt deposits are overconsolidated (OC) and exhibit $\sigma'_p$ greater than $\sigma'_{v0}$ (i.e., OCR greater than unity).

The $\sigma'_p$ corresponds to the effective vertical stress ($\sigma'_v$) beyond which major structural changes, including the breakdown of interparticle bonds (yielding), begin to occur. The selection of $\sigma'_p$ directly influences the determination of the compression index ($C_c$) and, if applicable, the recompression index ($C_r$) used for primary consolidation calculations. Consequently, accurate determination of $\sigma'_p$ is essential for reliable estimation of settlement.

\begin{figure}[H]
    \centering
    \includegraphics[width=0.5\linewidth]{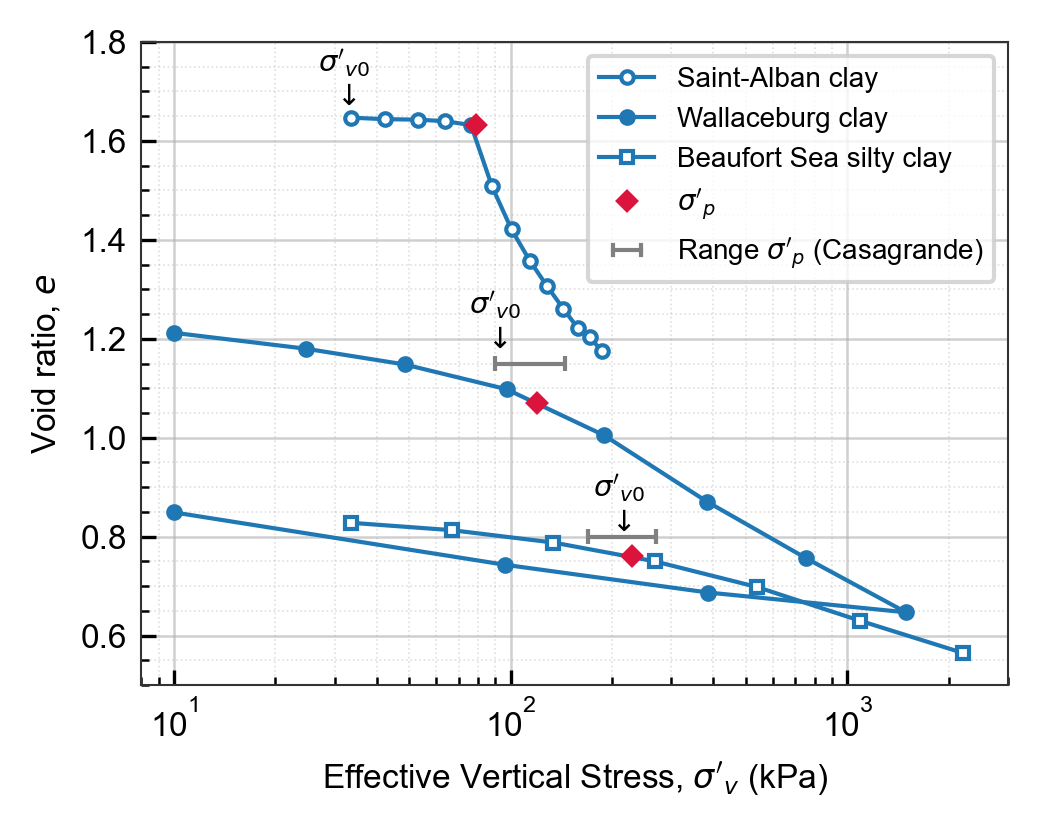}
    \caption{Consolidation $e$-log($P$) curves for Saint-Alban clay (after Terzaghi et al., 1996), Beaufort Sea silty clay, and Wallaceburg clay (after Becker et al., 1987).}
    \label{fig:fig1}
\end{figure}

\paragraph{Determination of \texorpdfstring{$\sigma'_p$}{sigma'p}}
The most widely used method for determining $\sigma'_p$ is the Casagrande (1936) graphical procedure \cite{casagrande1936determination}. The method is simple and widely accepted, but it may be  subjective because it relies on graphical interpretation to identify the point of maximum curvature. As a result, accurate determination of $\sigma'_p$ can be uncertain, particularly for silty clays and poor-quality specimens that exhibit rounded $e$-log($P$) curves. Figure~\ref{fig:fig1} presents the $e$-log($P$) curves of undisturbed sample of soft, sensitive Saint-Alban clay, Beaufort Sea silty clay, and Wallaceburg clay. The point of maximum curvature is clearly defined for the Saint-Alban clay but becomes less distinct for the more rounded Beaufort Sea silty clay and Wallaceburg clay’s. Since identifying the point of maximum curvature is the first step for determining $\sigma'_p$ using the Casagrande method, the range of possible $\sigma'_p$ corresponding to the identified curvature is compared with the reported $\sigma'_p$ for each specimen, as indicated by a red diamond on the respective curves.

\begin{figure}
    \centering
    \includegraphics[width=0.5\linewidth]{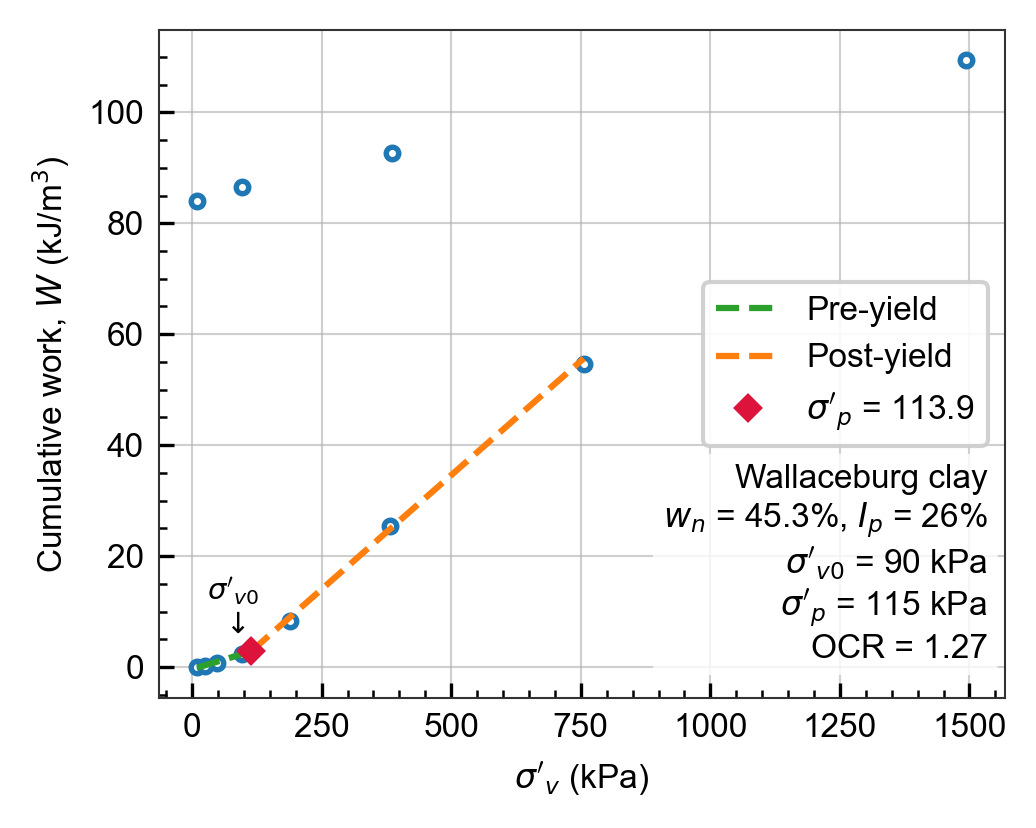}
    \caption{Work per unit volume interpretation for normally consolidated Wallaceburg clay estimated using \texttt{PyelogP} (after Becker et al., 1987). The pre-yield and post-yield lines intersect at $\sigma'_p$ of 113.9 kPa.}
    \label{fig:fig2}
\end{figure}

The strain-energy method \cite{Becker1987} uses the work per unit volume as a yield-criterion to identify the transition from small-strain to large strain response. Unlike the Casagrande graphical procedure, this method is based on the concept that the volumetric strain energy stored in a soil changes significantly when the applied stress exceeds the $\sigma'_p$. An important advantage of the method is its ability to estimate the $\sigma'_p$ for soils exhibiting rounded $e$-log($P$) curves and its applicability to both NC and OC clays \footnote{In Becker et al. (1987), the normally consolidated clays presented have OCR values between 1.0 and 1.3, while the overconsolidated clays have OCR values between 1.7 to 2.1.}. To determine the $\sigma'_p$ from an oedometer test, the cumulative work per unit volume is plotted against the applied effective vertical stress on an arithmetic linear scale. For NC clays, the pre-yield and post-yield portions of the curves can generally be represented by two approximately straight lines, with their intersection corresponding to $\sigma'_p$, as depicted in Figure~\ref{fig:fig2}. For OC clays, pre-yield response is represented by an approximately straight line up to $\sigma'_{v0}$, after which it bends into a distinct linear post-yield response. The intersection of the fitted lines represents the $\sigma'_p$.

\section{PyelogP Algorithm \& Methodology}

Because $\sigma'_p$ is determined from the intersection of two fitted straight lines, the procedure can be implemented in a computational algorithm. The remainder of this paper presents the core architecture of \texttt{PyelogP}, a Python library developed to automate the determination of $\sigma'_p$ from consolidation test data. The applicability of the library is demonstrated through case studies involving several well-documented clays reported in published literature.

Given a set of $e$-log($P$) data, \texttt{PyelogP} first computes the strain, work, and cumulative work corresponding to each data point, and subsequently removes any unload-reload cycles from the dataset. Only the data points within the recompression and virgin compression ranges are retained for determining the $\sigma'_p$ in the work-pressure (W-P) space. The $e$-log($P$) compression curve is then represented using a natural cubic spline, providing a smooth, piecewise-continuous approximation that captures localized nonlinear consolidation response while avoiding the oscillations and endpoint instability of high-order global polynomials. By maintaining $C^2$ continuity, which ensures continuity of the spline and its first and second derivatives across data points, the natural cubic spline provides smooth, continuous first (${d e}/{d\log \sigma'}$) and second (${d^2 e}/{d(\log \sigma')^2}$) derivatives without introducing jump discontinuities (e.g., \texttt{PCHIP}) or over-smoothing the yield transition (e.g., \texttt{UnivariateSpline}) across sparse consolidation datasets. \texttt{PyelogP} requires a minimum of six data points along the compression curve to ensure sufficient data density for the subsequent split-point linear regression.

The algorithm consists of two core components: split-point regression fitting in the W-P space and a thresholding procedure. 

\subsection{Split-point linear regression}

The W-P relationship shown in Figure~\ref{fig:fig2} can be formulated as a piecewise linear regression problem in which two straight lines are fitted to the data and joined at a common breakpoint (or “knot”), corresponding to the $\sigma'_p$ in our case. For a continuous piecewise linear model with a single breakpoint, the pre-yield and post-yield regions are represented by two linear segments share a common breakpoint. Let $c$ denote the breakpoint, $\beta_0$, $\beta_1$, $\beta_2$ denote the regression coefficients, and $\epsilon_i$ denote the residual error for observation $i$. The breakpoint is incorporated directly into the regression model through a continuity constraint, which forces the two linear segments to meet at $c$ while allowing the slope to change across the breakpoint. The truncated linear term, $x_i - c$, is defined as 0 when $x_i \leq c$ and $x_i - c$ when $x > c$. The resulting model can be expressed in piecewise form as: 

\begin{equation}
y_i =
\begin{cases}
\beta_0 + \beta_1 x_i + \varepsilon_i, & x_i \leq c, \\[6pt]
(\beta_0 - \beta_2 c) + (\beta_1 + \beta_2)x_i + \varepsilon_i, & x_i > c.
\end{cases}
\label{eq:piecewise_model}
\end{equation}

Although continuous piecewise linear regression can be optimized using goodness-of-fit measures such as the adjusted $R^2$ and the residual sum of squares, the statistically optimal breakpoint does not necessarily coincide with the physically meaningful $\sigma'_p$. Because the objective of this study is to determine $\sigma'_p$ in the W-P space, rather than enforcing continuity within a single piecewise regression model, the \texttt{PyelogP} adopts a different regression strategy by partitioning the dataset into pre-yield and post-yield subsets using a candidate breakpoint. For each candidate breakpoint, two independent linear regression models are fitted to the corresponding data subsets, and the mean squared errors (MSE) of the two regression models is computed. A search is then performed over all admissible candidate breakpoints to identify the optimal breakpoint, defined as the partition that minimizes the combined MSE of the two regression segments.

The \texttt{PyelogP} formulation is expressed as:

\begin{equation}
\hat{y}_i =
\begin{cases}
m_1 x_i + b_1, & x_i \leq k, \\
m_2 x_i + b_2, & x_i > k.
\end{cases}
\end{equation}

Where $m_1$ and $m_2$ are the independently fitted slopes, and $b_1$ and $b_2$ are the corresponding intercepts of the two regression lines. The variable $k$ represents the candidate breakpoint used to partition the dataset into pre-yield and post-yield subsets. In \texttt{PyelogP}, when the difference between $m_1$ and $m_2$ is less than $1 \times 10^{-4}$, meaning the regression lines are parallel, the algorithm stops. The optimal breakpoint is determined by minimizing the combined MSE of the two regression segments:

\begin{equation}
\min_{k \in K}
\frac{1}{n}
\left\{
\sum_{x_i \leq k} (y_i - \hat{y}_{1,i})^2
+
\sum_{x_i > k} (y_i - \hat{y}_{2,i})^2
\right\}
\end{equation}

where $K$ represents the set of admissible breakpoints satisfying the imposed physical constraints and $n = n_{pre-yield} + n_{post-yield}$. The optimal breakpoint defines the data partition used for regression fitting but is not directly taken as the estimated $\sigma'_p$. Instead, the final estimate of $\sigma'_p$ is obtained by calculating the intersection of the two regression lines fitted using the optimal partition.

Because the deformation response around the $\sigma'_p$ and post-yield region is inherently nonlinear, additional physical constraints (or thresholds) are incorporated to define the admissible breakpoint range. Thus, although MSE is used as the statistical criterion for selecting the optimal partition, the search is performed within a physically constrained domain. Conventional measures such as $R^2$ and MSE remain useful for evaluating statistical fit but are insufficient on their own to assess the physical validity of the estimated $\sigma'_p$. Unlike conventional continuous piecewise linear regression, in which the breakpoint is incorporated directly into a continuity-constrained regression model, \texttt{PyelogP} independently fits the two regression segments without imposing continuity and subsequently determines $\sigma'_p$ from their intersection.

\paragraph{Knee}
An initial candidate breakpoint is required in the \texttt{PyelogP} algorithm to separate the data. The point of maximum curvature ($\kappa_{max}$) can potentially be used to guide the selection of the initial candidate breakpoint, consistent with the Casagrande method. However, because curvature is a local, differential-geometry quantity based on the derivatives of a curve, the location is sensitive to data noise, dependent on scale, and difficult to identify when the transition is gradual. In simple terms, $\kappa_{max}$ corresponds to the location of the greatest geometric bending, or the sharpest bend. To overcome these limitations, the concept of a knee is introduced \cite{Salvador2004,Satopaa2011}.

\begin{equation}
\kappa(x) = \frac{f''(x)}{\left(1 + f'(x)^2\right)^{3/2}}
\end{equation}

\begin{figure}[t]
    \centering
    \begin{subfigure}[b]{0.48\columnwidth}
        \centering
        \includegraphics[width=\linewidth, height=6.5cm]{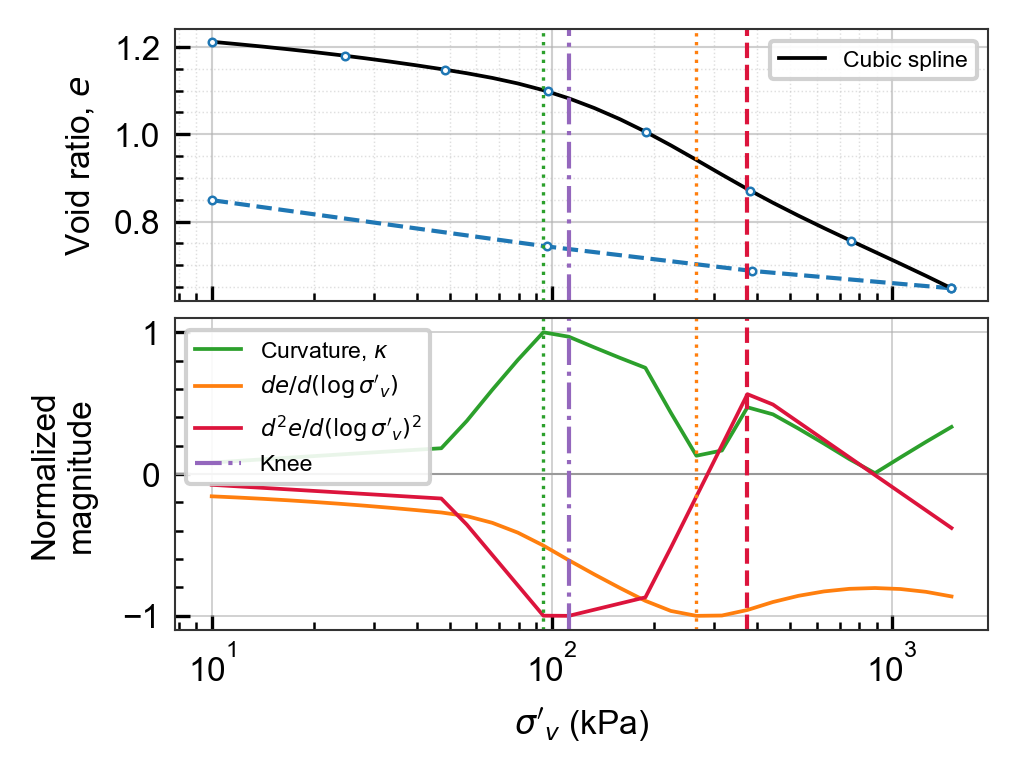}
    \end{subfigure}
    \hfill
    \begin{subfigure}[b]{0.48\columnwidth}
        \centering
        \includegraphics[width=\linewidth, height=6.5cm]{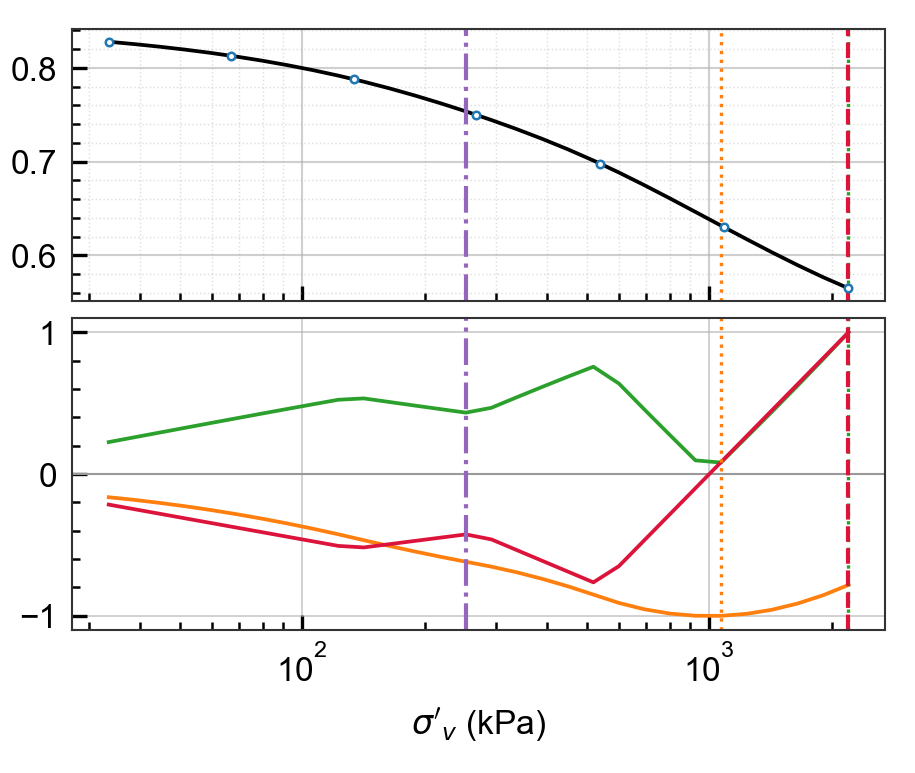}
    \end{subfigure}
    \caption{(Left) The locations of the point of maximum curvature ($\kappa_{max}$) and knee are close for the $e$-log($P$) curve of Wallaceburg clay. The maximum of second derivative (${d^2 e}/{d(\log \sigma')^2}$) is located at the flattening transition. (Right) The point of $\kappa_{max}$ overlaps with the maximum ${d^2 e}/{d(\log \sigma')^2}$ at the end of the curve for the Beaufort Sea silty clay.}
    \label{fig:fig3}
\end{figure}

A knee (or elbow) refers to a transition region in a curve where the rate of change shifts from one regime to another. In this context, it represents the point at which the slope changes appreciably, indicating a transition in the underlying response of the soil. In computer applications, knee points are commonly used for selecting the number of clusters in K-means algorithms and optimizing resource allocation in system design. Figure~\ref{fig:fig3} illustrates the first and second derivatives of the cubic spline, together with the corresponding curvature, the location of $\kappa_{\max}$, maximum ${d^2 e}/{d(\log \sigma')^2}$, and the identified knee points. A knee can occur at the same location as the $\kappa_{max}$, such as the $e$-log($P$) curve of structured Saint-Alban clay, or close to $\kappa_{max}$ when the maximum curvature is distinct. However, the locations of $\kappa_{max}$ and knee point are substantially different for the Beaufort Sea silty clay $e$-log($P$) curve. By using knee as the initial breakpoint to partition the data into pre-yield and post-yield subsets, a grid search can be performed over neighboring data points to identify the optimal regression segments. 

Because the knee represents the point on a curve where the overall response changes most noticeably, it provides a reliable initial estimate of breakpoint, $\sigma'_{v,knee}$. To compute $\sigma'_{v,knee}$, \texttt{PyelogP} integrates the Kneedle algorithm proposed by \cite{Satopaa2011} (via the open-source \texttt{kneed} Python library). The algorithm first normalizes the $e$-log($P$) compression curve into the unit domain of $[0, 1]^2$:

\begin{equation}
x = \frac{\log_{10}(\sigma'_i) - \min \log_{10}(\sigma')}
          {\max \log_{10}(\sigma') - \min \log_{10}(\sigma')},
\qquad
y = \frac{e_i - \min(e)}
          {\max(e) - \min(e)}.
\end{equation}

The knee point is then formally defined as the location of maximum perpendicular distance (or vertical deviation) from the normalized diagonal baseline, $L(x)=1-x$. In continuous terms, this corresponds to the point where the tangent to the normalized spline curve matches the slope of the secant line joining the curve boundaries.

\begin{figure}[H]
    \centering
    \includegraphics[width=0.4\linewidth]{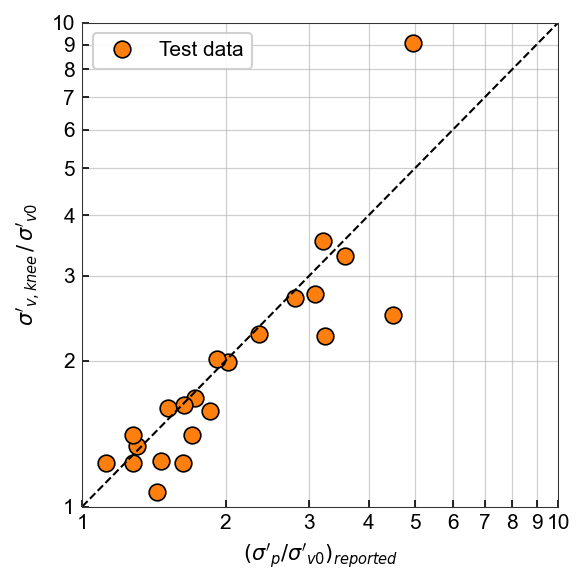}
    \caption{Comparison of $\sigma'_{v,knee}$ obtained through $e$-log($P$) curves and reported $\sigma'_p$ normalized to $\sigma'_{v0}$ using the test cases.}
    \label{fig:fig4}
\end{figure}

Figure~\ref{fig:fig4} presents $\sigma'_{v,knee}$ values compared with the reported $\sigma'_p$ normalized by $\sigma'_{v0}$ and plotted on a logarithmic scale to illustrate the distribution of the test data. The dataset consists of twenty-two $e$-log($P$) curves from published studies, representing slightly overconsolidated to highly overconsolidated clays with OCR values ranging from 1.1 to 5.0. Most of the data are from soft clays with OCR values below 3 because these soils are of primary interest in settlement analysis. As shown, a positive, linear relationship between $\sigma'_{v,knee}$ and reported $\sigma'_p$ values exists within the OCR range examined in this study. It should be noted that many of the reported $\sigma'_p$ values were determined using Casagrande graphical procedure, which produces a range of possible $\sigma'_p$ values as previously discussed. Therefore, some of the discrepancies may reflect uncertainty in the reported $\sigma'_p$ values rather than limitations of the knee or the proposed algorithm. However, the limited availability of $e$-log($P$) data for highly overconsolidated clays ($\mathrm{OCR} \geq 3$) precludes a more comprehensive evaluation of the algorithm in this range. Fortunately, most soft clay deposits are below OCR of 3.0. Accordingly, \texttt{PyelogP} adopts an OCR threshold of 3.0 for defining the pre-yield fitting region, as this value lies well within the range over which the knee-based estimate remains reliable, as discussed in the following section.

\subsection{Thresholding}

Although statistically based optimization identifies the optimal breakpoint in terms of statistical fit, the resulting breakpoint may not always represent the physically meaningful transition between pre-yield and post-yield response. To address this limitation, the \texttt{PyelogP} algorithm incorporates physically informed constraints for partitioning the pre-yield and post-yield data subsets. These thresholds are established based on the expected mechanical response of OC clays \cite{Becker1987} and the non-linear response within the virgin compression range. Specifically, the pre-yield segment is constrained by the $\sigma'_{v0}$, because stress states below this level are associated with recompression behavior rather than yield-related deformation. Similarly, the post-yield region is restricted to avoid excessive inclusion of highly nonlinear response associated with S-shaped virgin compression behavior. 

Within the pre-yield data subset, \texttt{PyelogP} introduces a cutoff at $\sigma'_{v0}$ for OC clays when the $\mathrm{OCR} \geq 3$. Since the optimal breakpoint is determined by minimizing the MSE, it represents the point at which a linear regression best fits the pre-yield data. Thus, when the deformation response transitions into yielding, the optimal breakpoint may extend beyond the linear portion into the region of gradual deviation. For clays with $\mathrm{OCR} < 3.0$, the entire pre-yield data subset is used for linear fitting. Accordingly, the algorithm selects the fitting range based on the current OCR value: for $\mathrm{OCR} \geq 3$, only data up to $\sigma'_{v0}$ are included in the pre-yield fit; otherwise, the full pre-yield subset is retained. Each iteration produces an updated $\sigma'_p$ from which a new OCR is calculated and compared with the threshold to determine the fitting range for the subsequent iteration. The process continues until either a maximum of five iterations is reached or the change in the estimated $\sigma'_p$ falls below a tolerance of $1 \times 10^{-4}$. 

While the pre-yield threshold is based on OCR, an additional criterion is required to define the extent of the post-yield fitting region because the virgin compression curve commonly exhibits increasing nonlinearity at higher stresses. As shown in Figure~\ref{fig:fig1}, the $e$-log($P$) curves exhibit a distinctly nonlinear response beyond the $\sigma'_p$. The maximum $C_c$ (${d e}/{d\log \sigma'}$) occurs immediately after $\sigma'_p$, after which the slope gradually decreases with increasing stress. This trend is most pronounced for the Saint-Alban clay, less evident for the Wallaceburg clay, where the transition occurs at approximately 400 kPa, and is not observed for the Beaufort Sea silty clay. The progressive flattening of the virgin compression curve has been attributed to the gradual destruction of natural soil structure, including particle bonding and fabric, during virgin loading. This behavior is more commonly observed in structured natural clays, such as sensitive marine clays, aged clays, and cemented clays. As destructuration progresses, the compression behavior increasingly approaches that of the corresponding reconstituted soil, resulting in a reduction in incremental compressibility at higher stress levels \cite{Burland1990, Leroueil1990}.

\texttt{PyelogP} defines the upper limit of the post-yield fitting region as the point where the virgin compression curve begins to transition from its maximum compressibility toward a progressively flatter response. On the $e$-log($P$) curve, this corresponds to the onset of decreasing compressibility, where the initially steep virgin compression slope recovers most rapidly. To identify this transition, \texttt{PyelogP} locates the maximum of the second derivative, ${d^2 e}/{d(\log \sigma')^2}$ of the spline-fitted $e$-log($P$) curve. The first derivative, ${d e}/{d\log \sigma'}$, represents the compression index, $C_c$, and its minimum corresponds to the maximum $C_c$ immediately following $\sigma'_p$ (see Figure~\ref{fig:fig3}). The maximum second derivative therefore marks the point at which the slope begins to flatten most rapidly, defining the upper limit of the post-yield fitting region. For the Wallaceburg clay, this transition occurs at 375 kPa, whereas the estimated $\sigma'_p$ is 111 kPa. Because \texttt{PyelogP} requires at least three data points to perform the post-yield linear regression, the next available data point beyond the transition is included when necessary.

\subsection{Software architecture and computational implementation}

\texttt{PyelogP} is implemented as an open-source, object-oriented Python package built on the standard scientific Python stack (\texttt{NumPy}, \texttt{SciPy}, and \texttt{kneed}). The package separates data preprocessing, spline generation, and split-point regression into modular functions. The preprocessing step automatically detects and strips unload-reload cycles, avoiding distortion during W-P transformation. The overall time complexity is $O(N^2)$, dominated by the windowed grid-search seeding step. Because per-curve point counts can range from under twenty up to roughly a thousand, at most, measured execution time stays in the low single-digit milliseconds per curve at the small end and remains under ~150 milliseconds per curve even at the largest curve sizes, while memory use scales linearly ($O(N)$) with the number of load increments, making batch processing of many test curves practical within an automated geotechnical pipeline.

\section{Validation \& Case Studies}

\begin{figure}[H]
    \centering
    \includegraphics[width=0.4\linewidth]{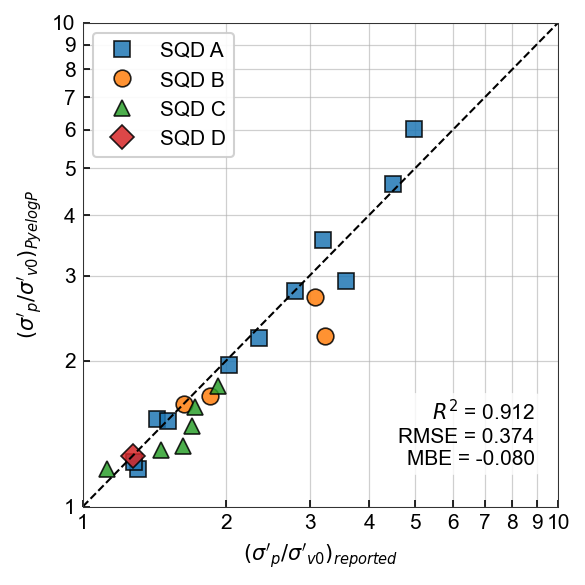}
    \caption{Comparison of \texttt{PyelogP (v0.2.0)} results.}
    \label{fig:fig5}
\end{figure}

A comparison of the proposed method using \texttt{PyelogP (v0.2.0)} (released in August 2026) against a series of validation test cases is presented in Figure~\ref{fig:fig5}. In addition to the $e$-log($P$) curves reported by \cite{Becker1987}, nineteen additional $e$-log($P$) datasets representing a wide range of clay deposits were used to evaluate the performance of \texttt{PyelogP}. These datasets include San Francisco Young Bay Mud \cite{Bonaparte1979}, Old Bay Clay \cite{Parks2019}, Boston Blue Clay \cite{Berman1993}, glacial till clays from Canada \cite{Soderman1970,MacDonald1970}, artificially sedimented clays \cite{Leonards1964}, $KCl$-saturated illitic mudrock \cite{Ge2019}, and structured marine clays from Louiseville, Canada; Mexico City, Mexico; and Changi, Singapore \cite{Terzaghi1996,Bo2015}. These test cases were selected because of their $e$-log($P$) shape, minimum number of data points, and provided consolidation parameters such as the initial void ratio ($e_0$), in situ $\sigma'_{v0}$, and reported $\sigma'_p$. As discussed previously, the limited availability of consolidation datasets with an $\mathrm{OCR} \geq 3$ restricts further evaluation of the \texttt{PyelogP} algorithm for highly overconsolidated soils. Nevertheless, the primary objective of the present study is settlement analysis for clays within the low to moderately overconsolidated range.

The test specimens were further classified according to the sample disturbance or Specimen Quality Designation (SQD) referenced in \cite{Terzaghi1996} and proposed by \cite{andresen1979ngi}, which is based on the volumetric strain criterion. SQD A represents high-quality, essentially undisturbed specimens characterized by less than 1\% volumetric strain ($\Delta \varepsilon_{vol}$) when the laboratory sample is subjected to a load equal to $\sigma'_{v0}$. In contrast, specimens classified as SQD C, with $\Delta \varepsilon_{vol}$ between 2 and 4\%, are generally considered relatively disturbed and rated as "Poor". Shelby tube and other thin-walled tube samples, which are commonly used for sampling soft clays, typically fall within this category. SQD A or B are generally recommended to determine consolidation parameter for settlement analysis, however, SQD C may be used when no other information is available.

As shown in Figure~\ref{fig:fig5}, the algorithm performs well even for lower-quality specimens, achieving an $R^2$ of 0.912, a root mean square error (RMSE) of 0.374, and a mean bias error (MBE) of -0.080. The RMSE of 0.374 represents a small prediction error relative to the normalized stress range between 1 and 5, while the high $R^2$ indicates that the computed values capture 91.2\% of the variability in the reported $\sigma'_p$ values. The computed $\sigma'_p$ values generally lie close to the 1:1 reference line, indicating good agreement with the reported values. However, the negative MBE indicates a slight systematic tendency of the algorithm to underestimate the reported $\sigma'_p$. A key observation is that the identified knee point of the $e$-log($P$) curve provides an effective initial point for separating the pre-yield and post-yield regions in the W-P space. Subsequent refinement of this initial estimate further improves the agreement, resulting in computed $\sigma'_p$ values that closely match the reported $\sigma'_p$.

\paragraph{Old Bay Clay}

Parks (2019) studied the engineering properties and geologic settings of the Old Bay Clay, also known locally as Yerba Buena Mud, of downtown San Francisco (SF). The Old Bay Clay is a Pleistocene estuarine or marine clay deposit that is widely encountered in the subsurface profile of the SF Bay Area and is of significant engineering interest due to its thickness and relevance to major construction projects. At the Transbay Transit Center site in downtown SF, the Old Bay Clay overlies the Alameda and Franciscan Formations, forming a relatively continuous layer approximately 24 m (80 ft) thick. The deposit is characterized as a dark greenish gray, stiff to hard, fat clay, consisting of 90 to 100\% fines, with natural water contents ($w_n$) ranging from 37 to 54\%, total unit weights from 16.5 to 18.4 $kN/m^3$ (105 to 117 pcf), liquid limits ($w_l$) from 60 to 68\%, and plasticity indices ($I_p$) from 37 to 44\%. Consolidation testing indicates that the Old Bay Clay is slightly overconsolidated to overconsolidated, with $\sigma'_p$ ranging from 620 to 880 kPa (13 to 18 ksf) and OCR values between 1.6 and 2.4. The $C_c$ ranges from 0.51 to 1.06 and $C_r$ from 0.01 to 0.07. 

\begin{figure}
    \centering
    \includegraphics[width=0.8\linewidth]{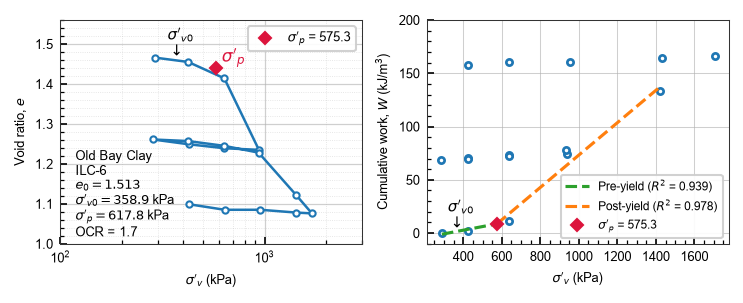}
    \caption{Old Bay Clay of downtown San Francisco (after Parks, 2019).}
    \label{fig:fig6}
\end{figure}

Shown on Figure~\ref{fig:fig6} are the $e$-log($P$) curve of incremental loading test sample “ILC-6” and its $\sigma'_p$ determined from \texttt{PyelogP}.  The test sample was obtained using epoxy-coated steel Shelby tube at a depth of approximately 35 m (115 ft). With an $e_0$ of 1.513 and an estimated $\sigma'_{v0}$ of 359 kPa (7.5 ksf), a $\Delta \varepsilon_{vol}$ of slightly less than 2.5\% (SQD C) is computed. Obtaining undisturbed samples from stiff or overconsolidated clays is generally difficult because stress relief during sampling often results in opening of fissures and potential softening that translates to additional sample disturbance (i.e., samples not representative of in situ conditions). Consequently, measured values of $C_r$ are particularly sensitive to sample disturbance. In practice, $C_r$ is commonly determined from the slope of the reloading portion of the $e$-log($P$) curve and compared to the $C_c$ over the stress range of between $\sigma'_p$ and approximately $2\sigma'_p$. Typical $C_r/C_c$ ratios range from 0.02 to 0.20 \cite{Terzaghi1996}. Higher ratios are generally associated with micaceous silts, fissured stiff clays, and shales, whereas lower ratios, or even lower, are characteristic of highly structured or cemented soft clay and silt deposits.

For sample ILC-6, the strain-energy method yielded a $\sigma'_p$ of 575.3 kPa, as shown on Figure~\ref{fig:fig6}, with an OCR of 1.6. Using the virgin compression line between $\sigma'_p$ and approximately $2\sigma'_p$, the corresponding $C_c$ is 0.91. The $C_r$, determined from the reloading cycle, is 0.06, resulting in a $C_r/C_c$ ratio of 0.07. The calculated indices are consistent with the published values, and the resulting ratio falls within the typical $C_r/C_c$ range for clays. 

\paragraph{Saint-Alban clay}

Tavenas, Leroueil and others \cite{Tavenas1974, Leroueil1978, Leroueil1983} conducted extensive investigations of the Saint-Alban clay deposit in Québec, Canada, as part of a series of laboratory and full-scale embankment studies that have become benchmark references for the behavior of sensitive marine clays. The Saint-Alban clay is a Holocene-aged Champlain Sea marine deposit that consists of a soft to firm, lightly overconsolidated clay with a structured fabric resulting from deposition, aging, and cementation. The deposit is characterized by $w_n$ generally ranging from 60 to 90\%, $w_l$ between approximately 40 and 60\%, $I_p$ of 15 to 30\%, and $e_0$ commonly between 1.5 and 2.5. The clay exhibits medium to high sensitivity, low porewater salinity due to post-depositional leaching, and OCR values generally between 1 and 2. Reported $C_c$ typically range from approximately 0.8 to 1.5, while $C_r$ range from about 0.05 to 0.20, reflecting the structured nature of the deposit and its response to unloading and reloading.

\begin{figure}[H]
    \centering
    \includegraphics[width=0.8\linewidth]{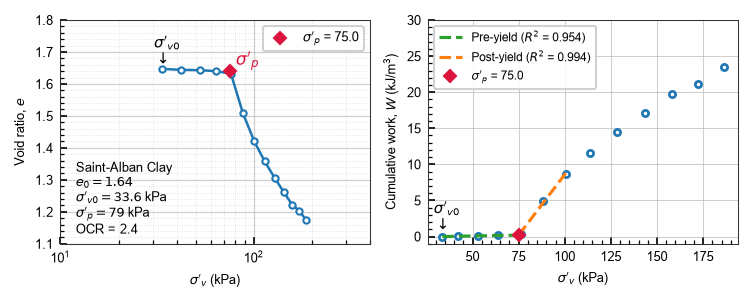}
    \caption{Saint-Alban clay (after Terzaghi et al., 1996).}
    \label{fig:fig7}
\end{figure}

Using the data from Saint-Alban $e$-log($P$) \cite{Terzaghi1996}, \texttt{PyelogP} calculates a $\sigma'_p$ of 75.0 kPa, corresponding to an OCR of 2.2. The computed $\Delta \varepsilon_{vol}$ in the recompression range is less than 1\% (SQD A). Using this $\sigma'_p$ value, $C_c$ is evaluated as 1.59 from $\sigma'_p$ to $2\sigma'_p$. Because of the high-quality sample, a $C_r$ value of 0.02 can be computed from the recompression range, resulting in a $C_r/C_c$ of 0.014 that is typical to structured clays. 

Overall, these case studies underscore the critical role of accurate $\sigma'_p$ determination in deriving reliable downstream consolidation parameters. 

\section{Conclusion, Recommendations \& Limitations}

This paper presents \texttt{PyelogP}, an open-source Python package that automates the determination of preconsolidation pressure ($\sigma'_p$) from oedometer test data using the strain-energy interpretation and split-point regression framework. By coupling natural cubic splines with automated knee point detection and physically constrained regression boundaries, \texttt{PyelogP} demonstrates promising performance based on the evaluated test datasets across various clay types and specimen quality. 

\paragraph{Recommendations and limitations}
\begin{itemize}
\item \textbf{Minor systematic underestimation of $\sigma'_p$}: Across several validation datasets, \texttt{PyelogP} slightly underestimates $\sigma'_p$ relative to the reported $\sigma'_p$ values. To execute split-point linear regression in W-P space, \texttt{PyelogP} requires at least six data points along the compression curve, which are partitioned into pre- and post-yield subsets. With additional thresholding, this partitioning often divides points unevenly, enforcing a three-point minimum constraint in the post-yield region can pull the regression line toward the pre-yield region (see Figure~\ref{fig:fig6} and ~\ref{fig:fig7}). Future releases will refine the post-yield criteria point selection rules.
\item \textbf{Dataset availability for high OCR clays}: \texttt{PyelogP} currently performs reliably on slightly to moderately overconsolidated clays ($OCR < 3$). Validation for highly overconsolidated clays ($OCR \geq 3$) remains limited due to the scarcity of complete, high-quality published consolidation datasets in this range. 
\item \textbf{Extension to CRS testing}: The current framework is optimized primarily for incremental loading (IL) tests alongside with limited constant rate-of-strain (CRS) datasets. Extension to continuous CRS test data and incorporating $\epsilon$-log($P$) represents an area for future release development.
\item \textbf{Incremental loading EOP}: For IL consolidation tests, ensuring that each loading step reaches End-of-Primary (EOP) conditions is essential. Terminating load step prematurely may alter the calculated void ratio, distorting the $e$-log($P$) curve and causing an underestimation of soil compressibility and $\sigma'_p$.  
\item \textbf{Visual data inspection and spline sensitivity to noise}: Because threshold boundary rely on cubic splines and their derivatives, minor testing artifacts, seating errors, or scatter can induce localized spline oscillations and false peaks. Users should visually inspect raw experimental data prior; minor adjustments or smoothing of noisy data points ensure the algorithm captures true physical soil response rather than measurement artifacts.
\end{itemize}

\paragraph{Code availability and package usage}
\texttt{PyelogP (v0.2.0)} is freely available as an open-source library on the Python Package Index (PyPI) under the BSD 3-Clause License and can be installed directly using \texttt{pip}:

\begin{verbatim}
pip install pyelogp
\end{verbatim}

The source code, example notebooks, and validation datasets can be accessed
via the public package repositories:

\begin{itemize}
    \item \textbf{PyPI distribution:}
    \url{https://pypi.org/project/pyelogp}

    \item \textbf{Source code and documentation:}
    \url{https://github.com/liangchow/PyelogP}
\end{itemize}

Researchers, practitioners, and developers are invited to explore the package, report feedback/issues, and contribute high-quality datasets to support ongoing open-source development and validation.

\section*{Acknowledgments}
The authors gratefully acknowledge Chunwei Ge for generously sharing experimental consolidation dataset for illitic mudrock A, and Zhongze Xu for testing and validating the \texttt{PyelogP} algorithm on a CRS test data. We also extend our sincere thanks to Margaret Parks for providing insightful comments regarding the Old Bay Clay case study in this manuscript.

\bibliographystyle{unsrt}  
\bibliography{references}

\end{document}